\documentclass[prl, aps,twocolumn,latexsym]{revtex4}

\usepackage{amssymb,color,ulem}
\usepackage{latexsym}
\usepackage{graphicx}
\usepackage{amsmath}
\usepackage{graphics}
\usepackage{cancel}
\usepackage{cleveref}

\begin{document}

\title{Landauer formula for interacting systems: a consistent non-perturbative approximation}

\author{Dan Klein}
\affiliation{Department of Condensed Matter Physics, Weizmann Institute of Science, Rehovot 76100, Israel}
\author{Karen Michaeli}
\affiliation{Department of Condensed Matter Physics, Weizmann Institute of Science, Rehovot 76100, Israel}

%\date{\today}

\begin{abstract}
Transport measurements are one of the most widely used methods of characterizing small systems in chemistry and physics. When interactions are negligible, the current through quantum dots, nanowires, molecular junctions, and other submicron structures can be obtained using the Landauer formula. Meir and Wingreen derived an exact expression for the current that also applies in the presence of interactions. This powerful theoretical tool requires knowledge of the exact Green's function. So far, an approximation extending beyond direct finite-order perturbation theory is missing. 
Here, we provide general expressions for both the electric and thermal currents where we expand the self-energy to the lowest order (frequently dubbed the GW approximation) but keep contributions to all orders in this quantity. Moreover, we show that the electric current is conserved only when the self-energy and vertex corrections are correctly included. We demonstrate that our formulae capture important non-perturbative features and hence, provide a powerful tool in cases where the exact solution cannot be found. 

\end{abstract}

\maketitle

Transport experiments are arguably the most common method to probe the electronic properties of nanoscopic or microscopic systems. Over the years, such measurements revealed various fascinating phenomena unique to the submicron world. Examples include conductance steps in clean systems~\cite{Foxon,Jones}, universal conductance fluctuations~\cite{Lee}, and the Coulomb blockade~\cite{Meirav,Beenakker}. Extensive theoretical studies accompanied the experimental effort, and the  Landauer formula~\cite{Landauer,Buttiker,Imry} played a central role in understanding many of the observations. Landauer's expression describes the current flowing through a system connected to leads---electron baths with a well-defined temperature and chemical potential. Specifically, the electric~\cite{Landauer} and thermal~\cite{SivanImry} currents are formulated in terms of the scattering probabilities through the finite region.

Explaining the observation of conductance steps in ballistic nanowires~\cite{Foxon,Jones} and predicting the signature of Aharonov Bohm oscillations in small metallic rings~\cite{Gefen,Webb} are among the many applications of this powerful method.   By contrast, the Landauer formula cannot capture the physics behind the appearance of Coulomb blockade oscillations~\cite{Beenakker2}, the long-range electron transfer in molecular chains~\cite{NitzanBook}, and various other effects that stem from interactions~\cite{McEuen}. Meir and Wingreen~\cite{Meir} have generalized Landauer's expression to interacting systems. The many-body Green's function (GF) replaces the scattering probability in their formulation. The Meir-Wingreen formula (MWF) has successfully explained the transport properties of quantum dots~\cite{Meir2} and molecular chains with a few sites~\cite{Richter}. The small Hilbert space in these cases facilitates obtaining the exact many-body GF. In larger systems, the current is commonly calculated using a perturbative expansion in the interaction~\cite{Galperin,Aharony,Vignale,Solomon}. Going beyond perturbation theory still remains a challenge~\cite{Kamenev,Segal1,Segal2}.

Various techniques were developed to outperform the straightforward perturbation expansion in bulk systems. For example, in the random phase approximation (RPA), we calculate the screened potential self-consistently~\cite{RPA,Mahan}  using the lowest order contribution to the polarization operator $\Pi$. %$V_{sc}$ self-consistently using the lowest order approximation to the polarization operator $\Pi$, i.e., $V_{sc}=V_{0}+V_0\Pi V_{sc}$. 
The RPA amounts to summing over an infinite subset of contributions to the screened potential---we include corrections to all orders in $\Pi$, which itself is found within perturbation theory.  A similar procedure is used to derive disorder corrections to the electronic GF within the self-consistent Born approximation~\cite{RammerBook}. It would be tempting -- but dangerous -- to simply insert such approximate Green's functions into the MWF. The resulting expression for the current misses crucial contributions and may not conserve charge.  

\begin{figure}[t]
       \includegraphics[width=0.15\textwidth]{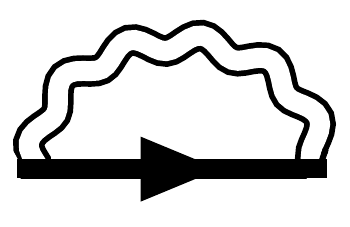}
                \caption{\textbf{Illustration of the lowest order approximation of the self-energy given by Eq.~\eqref{Self-Energy}.} The self-energy consists of one bare (non-interacting) electronic GF (single line) and one propagator of the boson mode (double wavy line). The latter is given by~\eqref{PhononPropagator} for interactions with large baths of bosons, and by Eq.~(S13) in the Supplementary Material for electron-electron interactions or small baths.  }
\label{fig.1}
\end{figure}

In this work, we find expressions for the electric and thermal currents that contain all contributions with the lowest order correction to the electronic self-energy (see Fig.~\ref{fig.1}).  An essential feature of our formula is that we include all self-energy and vertex corrections at the same level of approximation. The main part of the paper focuses on electron-phonon interactions as a representative example of coupling between the electronic system and any bosonic modes. The modifications required for including electron-electron interactions are given in the Supplementary Material. To demonstrate the strength of our result, we perform a calculation of phonon-induced charge transfer through a molecular bridge. Within perturbation theory, a non-zero current first arises at orders equal to the lattice size. By contrast, our expression instantly captures the formation of a polaronic band.  
%\david{at the xth order?}   

Performing a perturbative expansion of the self-energy and not the GF is at the heart of the GW approximation~\cite{Hedin,Louie,Gunnarsson}. Such an approach is typically used to study the effect of electron-electron interactions in molecules and solids. Although the expansion of the self-energy to the lowest order is not always justified, it predicts the spectrum of solids~\cite{Sham1,Sham2,Pollmann,Louie2,Louie3,Wilkins,Millis,Rinke} very accurately.  Similarly, the GW approximation captures qualitative behavior and scaling with the temperature of many phenomena induced by electron-phonon coupling~\cite{Cohen,Kresse}. We provide a simple scheme for implementing this powerful method in calculating transport coefficients.

Following Ref.~\cite{Meir}, we study a finite system of interacting electrons connected to two infinite, non-interacting leads
\begin{align}\label{Hamiltonian}
	H=&H_{\text{sub}}+\sum_{\vec{\nu}_{\text{L/R}}}\left[\epsilon_{\vec{\nu}_{\text{L}}}d_{\vec{\nu}_{\text{L}}}^{\dag}d_{\vec{\nu}_{\text{L}}}+\epsilon_{\vec{\nu}_{\text{R}}}d_{\vec{\nu}_{\text{R}}}^{\dag}d_{\vec{\nu}_{\text{R}}}\right]\\
	+& \sum_{\vec{n},\vec{m},\vec{\nu}_{\text{L/R}}}\left[\gamma_{\vec{n},\vec{m}}^{\vec{\nu}_{\text{L}}}c_{\vec{n},\vec{m}}^{\dag}d_{\vec{\nu}_{\text{L}}}\hspace{-0.6mm}+\gamma_{\vec{n},\vec{m}}^{\vec{\nu}_{\text{R}}}c_{\vec{n},\vec{m}}^{\dag}d_{\vec{\nu}_{\text{R}}}\hspace{-0.6mm}+\text{H.c}.\right].\nonumber
\end{align}
Here, $c_{\vec{n},\vec{m}}^{\dag}$ creates an electron with state $\vec{m}$ on site  $\vec{n}$ inside the finite subsystem. The index $\vec{m}$ contains all quantum numbers characterizing the electronic state such as spin and orbital. Similarly,  $d_{\vec{\nu}_{\text{L}}}^{\dag}$ ($d_{\vec{\nu}_{\text{R}}}^{\dag}$) creates an electron at a state $\vec{\nu}_{\text{L}}$ ($\vec{\nu}_{\text{R}}$) in the left (right) lead. Both leads are at equilibrium with temperature $T_{\text{L/R}}$ and chemical potential $\mu_{\text{L/R}}$. Such a description applies to systems of any dimension. The information on the leads' positions is encoded in the coupling parameter $\gamma_{\vec{n},\vec{m}}^{\vec{\nu}_{\text{L/R}}}$.  The Hamiltonian $H_{\text{sub}}$ describes the electrons inside the finite subsystem and their interactions with each other or the environment. The latter contains all modes not coupled directly to the leads, for instance, phonons.

\begin{figure}[t]
       \includegraphics[width=0.45\textwidth]{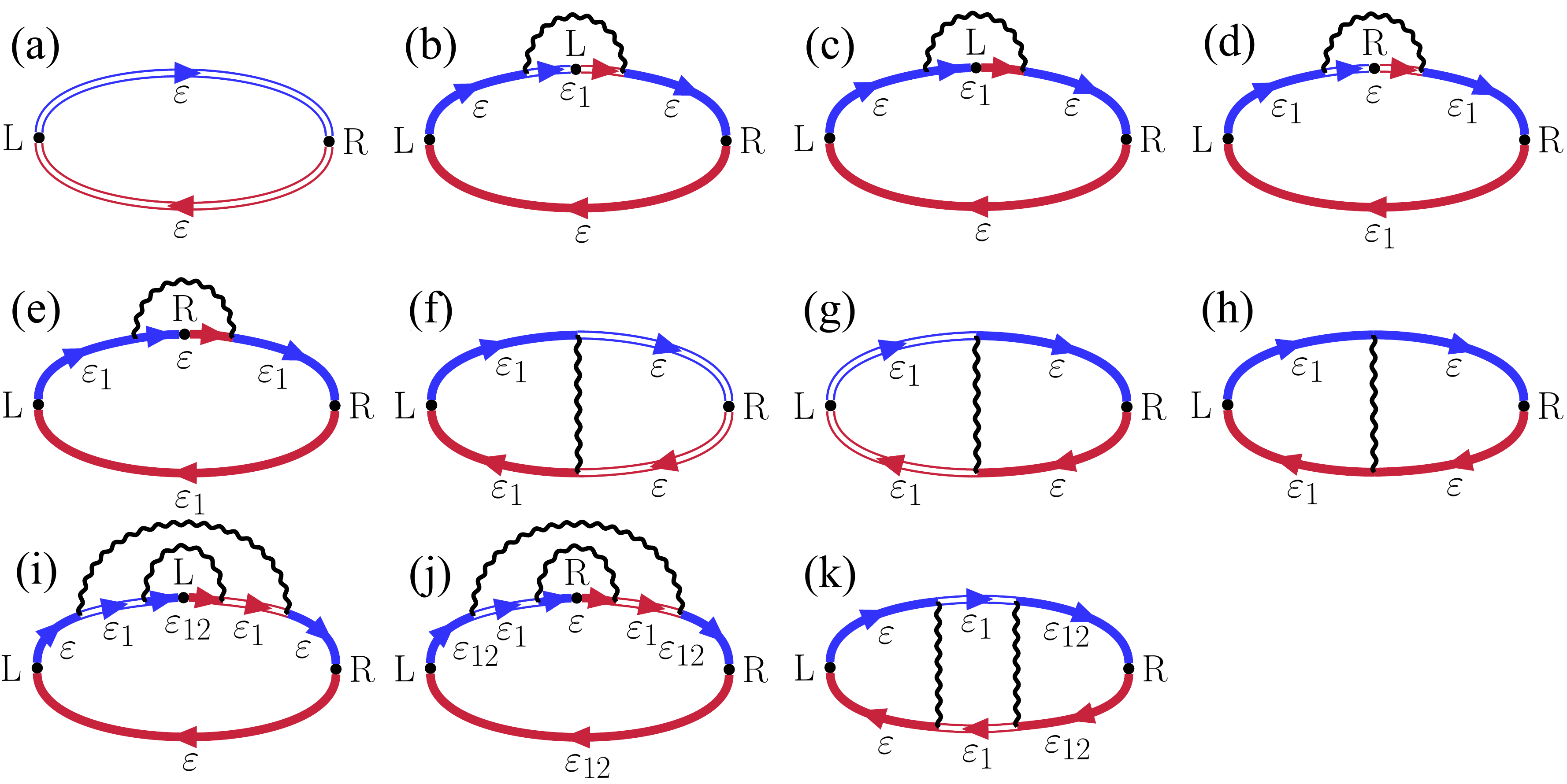}
                \caption{\textbf{Diagrammatic representation of contributions to the currents given by Eqs.~\eqref{CurrentDiagaram} and~\eqref{2CurrentDiagaram}}. The single and double lines are the bare and dressed electronic Green's function, respectively [see Eq.~\eqref{GRA}]. We use the color code to indicate the retarded (blue) and advanced (red) GFs. The indices L and R denote the lead from which the GFs end or start, and $\varepsilon_1=\varepsilon-\omega_1$ while $\varepsilon_{12}=\varepsilon-\omega_1-\omega_2$. The single wavy line represents the combination $D^{R}-D^{A}$, in contrast to the double line entering the self-energy (Fig.~\ref{fig.1}). }
\label{fig.2}
\end{figure}

The electric current flowing from the subsystem to the $j$-th lead ($j=\text{L/R}$) is given by the MWF
\begin{align}\label{MW}
J_{j}=\frac{ie}{\hbar}\int{\text{d}\varepsilon}\sum_{\mathbf{n},\mathbf{n}'}\Gamma_{\mathbf{n}',\mathbf{n}}^{j}&\left\{G_{\mathbf{n},\mathbf{n}'}^{<}(\varepsilon)[1-f_j(\varepsilon)]\right.\\\nonumber&\hspace{10mm}\left.+G_{\mathbf{n},\mathbf{n}'}^{>}(\varepsilon)f_j(\varepsilon)\right\}.
\end{align}
Here $G_{\mathbf{n}',\mathbf{n}}$ is the fully dressed Keldysh GF~\cite{Rammer}, and  $<$ and $>$ denote its lesser and greater components.  For simplicity, we combine all indices into a single one, $\mathbf{n}={\vec{n},\vec{m}}$, and hence the Keldysh GF can be written in terms of the single electron operators as  $G_{\mathbf{n},\mathbf{n}'}(t)=-i\left< T_{c}\left[ c_{\vec{n},\vec{m}}(t) c_{\vec{n}',\vec{m}'}^{\dag}(0)\right]\right>$.  The GF entering the MWF is a function of the frequency, i.e., the Fourier transform of $G_{\mathbf{n},\mathbf{n}'}(t)$.  The bare (non-interacting) current vertex is $\Gamma_{\mathbf{n}',\mathbf{n}}^{j}=2\pi \rho_{\vec{\nu}}^{j}P_{\vec{m}',\vec{m}}^{\vec{\nu}}\delta_{n,j}\delta_{n',j}$, and $f_{j}(\varepsilon)$ is the Fermi-Dirac distribution function of the same lead. The parameter $\rho_{\vec{\nu}}^{j}$ is the density of states in the lead $j$ per quantum number $\vec{\nu}$, while $P_{\vec{m}',\vec{m}}^{\vec{\nu}}$ transforms the basis of states in the lead to that of the subsystem. The coupling to the leads as well as the interactions renormalize the lesser GF 
\begin{align}
&G^{<}=i\sum_{j=L,R}f_jG^{R}\cdot\Gamma^{j}\cdot G^{A}+G^{R}\cdot\Sigma^{<}\cdot G^{A}.\label{G<>}
\end{align}
Here and below, we use dots to denote the product of matrices, i.e., $[A\cdot B]_{\mathbf{n},\mathbf{n}'}^{\varepsilon}=\sum_{\mathbf{p}}A_{\mathbf{n},\mathbf{p}}(\varepsilon)B_{\mathbf{p},\mathbf{n}'}(\varepsilon)$. We replace the lesser index with the greater one and the distribution function by $-1+f_j$  to obtain  $G^{>}(\varepsilon)$.  The retarded and advanced components of the GF are 
\begin{align}\label{GRA}\nonumber
\left[G^{R,A}\right]^{-1}&=\left[\varepsilon-H_{\text{sub}}^{\text{el}}\right]^{-1}\pm\frac{i}{2}\left[\Gamma^{\text{L}}+\Gamma^{\text{R}}\right]-\Sigma^{R,A}\\
&\equiv g^{-1}-\Sigma^{R,A}.
\end{align}
The Hamiltonian $H_{\text{sub}}^{\text{el}}$ describes the electrons of the finite subsystem alone in the absence of interactions. We define $g$ to be the electron GF in the subsystem, including only the effects of the leads, while the self-energy $\Sigma(\varepsilon)$ is fully dressed with interactions. We include only the lowest order corrections to the self-energy $\Sigma^{(1)}$, i.e., apply the GW approximation.

Substitution of the self-energy in Eqs.~\eqref{MW}-\eqref{GRA} with the approximate one, $\Sigma\rightarrow\Sigma^{(1)}$, generically results in a spurious violation of current conservation. This difficulty becomes apparent in systems with sufficiently complex dynamics and has therefore been mostly overlooked. By contrast, the exact Landauer formula and the MWF clearly satisfy $J_{\text{L}}= -J_{\text{R}}$~\cite{Ness}. Thus, such a simple substitution is not a consistent approximation to the current. We note that the self-energy consists of a correction to the single-particle spectrum and a relaxation rate. The latter is a manifestation of electronic transitions between different states by scattering events. Vertex corrections account for the current carried by the scattered electrons, thereby maintaining charge conservation. Following B\"{u}ttiker's idea~\cite{Bprobes}, the effect of interactions is often studied by attaching additional probes into the system and calculating the current from the Landauer formula. The extra leads introduce scattering rates into the electronic Green's functions. Charge conservation is guaranteed when the chemical potentials of the B\"{u}ttiker probes are chosen so that there is no current flowing from them to the system.  Fine-tuning the chemical potentials has the same role as correctly including the self-energy and vertex contributions. Self-consistent calculations of the chemical potentials are straightforward in linear response but become increasingly challenging far from equilibrium.

 The above discussion illustrates the delicate balance between self-energy and vertex corrections.  Furthermore, we see that the condition $J_{\text{L}}+J_{\text{R}}=0$ can help detect inconsistent approximations to the current. In the Green's function formalism, charge conservation is encoded in the Dyson equation, which we express in a more revealing form as
\begin{align}\label{WardIdentity}
&G^{R}-G^{A}=G^{R,A}\hspace{-1mm}\cdot\hspace{-1mm}\left[-i\Gamma^{L}-i\Gamma^{R}+\Sigma^{R}-\Sigma^{A}\right]\hspace{-1mm}\cdot\hspace{-1mm}G^{A,R}.
\end{align}
Up to here, our derivation applies for any subsystem. Next, we use electron-phonon coupling as a prototype for interactions between electrons and any bosonic mode. It can be applied, for example, for electrons in a molecular junction where vibrations play an important role in charge transfer or a quantum dot coupled to an optical cavity. The Hamiltonian of the finite subsystem $H_{\text{sub}}=H_{\text{el}}+H_{\text{ph}}+H_{\text{el-ph}}$ is a sum of the non-interacting electron and phonon contributions as well as the coupling between them~\cite{Frohlich}
\begin{equation}\label{el-ph}
H_{\text{el-ph}}=\sum_{\vec{n},\vec{n}',\vec{m},\vec{m}',\vec{k}}V_{\vec{n}',\vec{m}';\vec{n},\vec{m}}^{\vec{k}}(b_{\vec{k}}+b_{\vec{k}}^{\dag})c_{\vec{n}',\vec{m}'}^{\dag}c_{\vec{n},\vec{m}}.
\end{equation} 
Here, $b_{\vec{k}}^{\dag}$ creates a phonon excitation with frequency $\omega_{\vec{k}}$, and the interaction vertex is generically non-diagonal in space or $\vec{m}$ indices. At lowest order in the interaction, the components of the self-energy are
\begin{subequations}\label{Self-Energy}
\begin{flalign}\label{Self-Energy<,>}
\Sigma_{\mathbf{n},\mathbf{n}'}^{<,>}(\varepsilon)\hspace{-1mm}=\hspace{-1mm}\sum_{\vec{k},\mathbf{p},\mathbf{p}'}\int\hspace{-1mm}\frac{\text{d}\omega}{2\pi}D_{\mathbf{n},\mathbf{p};\mathbf{p}',\mathbf{n}'}^{<,>}(\vec{k},\omega)g_{\mathbf{p},\mathbf{p}'}^{<,>}(\varepsilon-\omega).
&&
\end{flalign}
\vspace{-5mm}
\begin{flalign}\label{Self-EnergyRA}\nonumber
\Sigma_{\mathbf{n},\mathbf{n}'}^{R,A}(\varepsilon)&=
\sum_{\vec{k},\mathbf{p},\mathbf{p}'}\int\hspace{-1mm}\frac{\text{d}\omega}{2\pi}\left[D_{\mathbf{n},\mathbf{p};\mathbf{p}',\mathbf{n}'}^{>}(\vec{k},\omega)g_{\mathbf{p},\mathbf{p}'}^{R,A}(\varepsilon-\omega)\right.\\
&\left.+D_{\mathbf{n},\mathbf{p};\mathbf{p}',\mathbf{n}'}^{R,A}(\vec{k},\omega)g_{\mathbf{p},\mathbf{p}'}^{<}(\varepsilon-\omega)\right].
&&
\end{flalign}
\end{subequations}
We assume a sizeable bosonic bath and therefore neglect the renormalization of the phonon modes. Thus, the phonon propagator maintains a simple form 
\begin{align}\label{PhononPropagator}
D_{\mathbf{n},\mathbf{p};\mathbf{p}',\mathbf{n}'}^{R,A}(\vec{k},\omega)=\frac{V_{\mathbf{n},\mathbf{p}}^{\vec{k}}D_{\vec{k}}^0V_{\mathbf{p}',\mathbf{n}'}^{\vec{k}}}{\omega-\omega_{\vec{k}}\pm i\delta}-\frac{V_{\mathbf{n},\mathbf{p}}^{\vec{k}}D_{\vec{k}}^0V_{\mathbf{p}',\mathbf{n}'}^{\vec{k}}}{\omega+\omega_{\vec{k}}\pm i\delta}.
\end{align}
The lesser and greater components are $D^{<}(\omega)=N^{\text{ph}}(\omega)[D^{R}(\omega)-D^{A}(\omega)]$ and $D^{>}(\omega)=[1+N^{\text{ph}}(\omega)][D^{R}(\omega)-D^{A}(\omega)]$, respectively; $N^{\text{ph}}(\omega)$ is the Bose-Einstein distribution with the phonon temperature  $T_{\text{ph}}$. In the Supplementary Material, we include corrections to the bosonic modes within the RPA and derive the current in the presence of electron-electron interactions.

We first  consider equal temperatures $T_{\text{L/R}}=T_{\text{ph}}$, and derive the electric current flowing from the interacting subsystem to the left lead as a response to an applied voltage $V=(\mu_{\text{L}}-\mu_{\text{R}})/e$. The current is written using its diagrammatic representation (Fig.~\ref{fig.2})
\begin{align}\label{CurrentDiagaram}
&J_{\text{L}}=-\frac{e}{\hbar}\int \text{d}\varepsilon\Re\mathcal{I}_a\left[f_{\text{L}}^{\varepsilon}-f_{\text{R}}^{\varepsilon}\right]\\\nonumber
&+\frac{e}{2\hbar}\int\frac{\text{d}\varepsilon \text{d}\omega_1}{(2\pi)}\Re\left[\mathcal{I}_b-\mathcal{I}_c+\mathcal{I}_d-\mathcal{I}_e+2i\mathcal{I}_f+2i\mathcal{I}_g-2i\mathcal{I}_h\right]\\\nonumber
&\times
\left[f_{\text{R}}^{\varepsilon}-f_{\text{L}}^{\varepsilon-\omega_1}\right]\left[N_{\omega_1}^{\text{ph}}-N_{\omega_1+eV}^{\text{ph}}\right]\\\nonumber
&
+\frac{ie}{2\hbar}\int\frac{\text{d}\varepsilon \text{d}\Omega\text{d}\omega_1\text{d}\omega_2}{(2\pi)^2}\delta(\Omega-\omega_1-\omega_2)\Re\left[\mathcal{I}_i+\mathcal{I}_j+2i\mathcal{I}_k\right]\\\nonumber
&\times
\left[f_{\text{R}}^{\varepsilon}-f_{\text{L}}^{\varepsilon-\Omega}\right]
\left[N_{\omega_2}^{\text{ph}}-N_{-\omega_1}^{\text{ph}}\right]\left[N_{\Omega}^{\text{ph}}-N_{\Omega+eV}^{\text{ph}}\right].
\end{align}
Here, $\mathcal{I}_\alpha$ refers to the product of electron and phonon GFs as indicated by diagram $\alpha$. The voltage enters the equation through the Fermi-Dirac distribution functions of the two leads $f_{j}(\varepsilon)=[e^{(\varepsilon-\mu_j)/T_{j}}+1]^{-1}$. The diagrams include the bare (single line) and dressed (double line) electronic GFs. Notice that the phonon propagators explicitly shown in the diagrams (single lines) are different from those entering the self-energy (double line). The former denotes the combination  $D^{R}-D^{A}\propto\Im D^{R}$, while the latter [Eq.~\eqref{PhononPropagator}] also includes the real part. The full expressions for each diagram are given in the Supplementary Material. The current flowing from the subsystem to the right lead $J_{\text{R}}$ is obtained by interchanging the left and right leads everywhere in Eq.~\eqref{CurrentDiagaram}. Crucially, the currents satisfy $J_{\text{L}}+J_{\text{R}}=0$  (see Supplementary Material).

\begin{figure}[t]
       \includegraphics[width=0.35\textwidth]{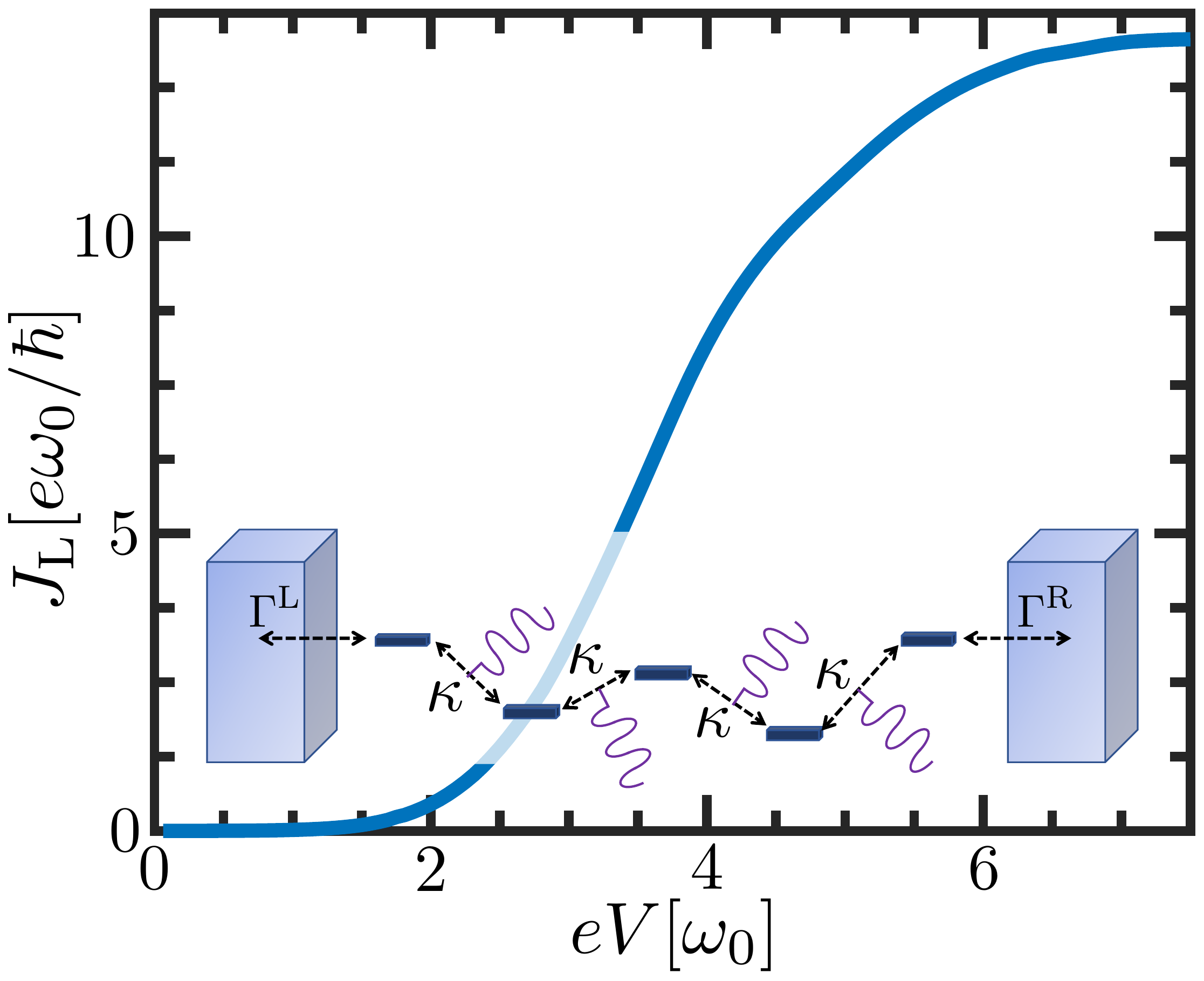}
                \caption{\textbf{Phonon-assisted current-voltage characteristic}. The current as function of $eV=\mu_{L}-\mu_{R}$ is calculated using~\eqref{CurrentDiagaram} for the Hamiltonian in Eq.~\eqref{Example}. This Hamiltonian is a toy model for molecular chains or systems in the small polaron limit. Our non-perturbative expression for the current captures the formation of an electronic band of a non-zero width as a consequence of electron-phonon interactions. }
\label{fig.4}
\end{figure}

We demonstrate the strength of our expression with an example of transport through a disordered wire assisted by optical phonons 
\begin{align}\label{Example}
H_{\text{sub}}&=\sum_{n=1}^{N}\epsilon_{n}c_{n}^{\dag}c_{n}+\omega_0\sum_{n=1}^{N}b_{n}^{\dag}b_n\\\nonumber
&+\kappa\sum_{n=1}^{N-1}(b_n-b_n^{\dag}+b_{n+1}-b_{n+1}^{\dag})\left[c_{n}^{\dag}c_{n+1}+\text{h.c}\right].
\end{align}
The on-site energies in the first term are randomly drawn from a uniform distribution in the domain $\epsilon_n\in [-W,W]$. This model describes charge transfer through a molecular bridge~\cite{Marcus,NitzanBook} or a one-dimensional system in the small polaron limit~\cite{Holstein,Emin}. Importantly, no current can flow in the absence of interactions. Within perturbation theory in the interaction, we must expand to an order equal or higher than the lattice length ($N=5$ in our example) to obtain a non-zero current. High order is needed because emission or absorption of phonons accompanies each hopping event. Our expression for the current given by Eq.~\eqref{CurrentDiagaram} consists of corrections to all orders in $\Sigma^{(1)}$. Consequently, it captures the formation of a conduction band, evident from the sigmoidal curve, when straightforwardly applied to the model Hamiltonian in Eq.~\eqref{Example}. The current is shown in Fig.~\ref{fig.4} for  $W=\omega_0$, $\Gamma^{\text{L}/\text{R}}=\kappa=T_{\text{ph}}=0.25\omega_0$, and  $\mu_\text{L}=-3.75\omega_0$. This simple example already illustrates the power of our derivation and its non-perturbative nature.

\begin{figure}[t]
       \includegraphics[width=0.45\textwidth]{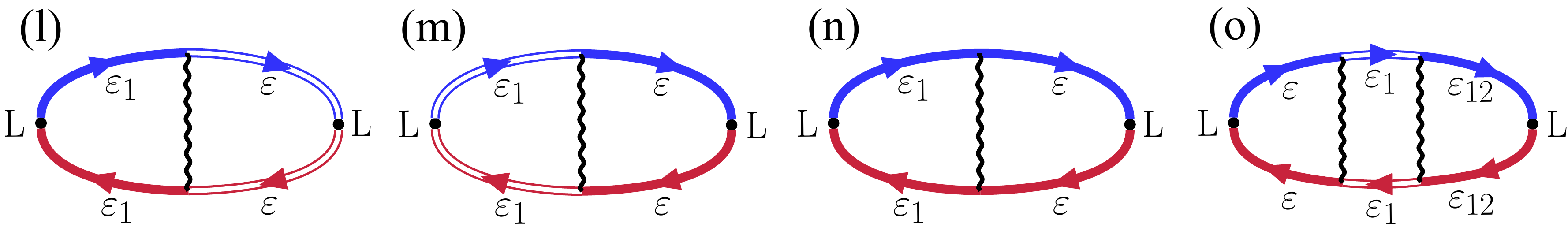}
                \caption{\textbf{Diagrammatic representation of contributions appearing in the general thermoelectric response}.  These diagrams, in addition to those given in Fig.~\ref{fig.2}, enter the expression for the electric and heat currents given by Eq. ~\eqref{2CurrentDiagaram}.}
\label{fig.3}
\end{figure}

We turn now to derive the electric and thermal currents induced by both an applied voltage and a temperature difference between the two leads. 
We find several new processes affecting the thermoelectric and thermal responses that are absent from Eq.~\eqref{CurrentDiagaram}. The diagrammatic representations of these additional contributions are shown  in Fig.~\ref{fig.3}. Together with the terms in Fig.~\ref{fig.3},  the currents can be written as   
\begin{align}\label{2CurrentDiagaram}
&J_{\text{L}}^{e,th}=\frac{1}{\hbar}\int{\text{d}\varepsilon}\Re\mathcal{A}_1+\frac{1}{2\hbar}\int\frac{\text{d}\varepsilon \text{d}\omega_1}{2\pi}\Re\mathcal{A}_2\\\nonumber
&+\frac{i}{2\hbar}\int\frac{\text{d}\varepsilon \text{d}\Omega \text{d}\omega_1 \text{d}\omega_2}{(2\pi)^2}\left[N_{\omega_2}^{\text{ph}}-N_{-\omega_1}^{\text{ph}}\right]
\delta(\Omega-\omega_1-\omega_2)\Re\mathcal{A}_3,
\end{align}
with
\begin{subequations}\label{2CurrentDiagaram-Elements}
\begin{flalign}\label{2CurrentDiagaram-Elements1}
&\mathcal{A}_1=\eta_{\varepsilon}\mathcal{I}_a\left[f_{\text{L}}^{\varepsilon}-f_{\text{R}}^{\varepsilon}\right];&&
\end{flalign}
\vspace{-7mm}
\begin{flalign}\label{2CurrentDiagaram-Elements2}\nonumber
&\mathcal{A}_2=\left[f_{\text{R}}^{\varepsilon}\hspace{-0.5mm}-\hspace{-0.5mm}f_{\text{R}}^{\varepsilon-\omega_1}\right]\left[N_{\omega_1}^{\text{ph}}-N_{\omega_1}^{\text{R}}\right]\eta_{\varepsilon-\omega_1}\left[\mathcal{I}_d-\mathcal{I}_e\right]\\\nonumber
&-\left[f_{\text{L}}^{\varepsilon}\hspace{-0.5mm}-\hspace{-0.5mm}f_{\text{L}}^{\varepsilon-\omega_1}\right]
\left[N_{\omega_1}^{\text{ph}}\hspace{-0.5mm}-\hspace{-0.5mm}N_{\omega_1}^{\text{L}}\right]
\left\{
2i\eta_{\varepsilon-\omega_1}\left[\mathcal{I}_l+\mathcal{I}_m-\mathcal{I}_n\right]\right.\\
&\left.-\eta_{\varepsilon}\left[\mathcal{I}_b-\mathcal{I}_c\right]\right\}
-\left[f_{\text{R}}^{\varepsilon}-f_{\text{L}}^{\varepsilon-\omega_1}\right]\left[N_{\omega_1}^{\text{ph}}-N_{\omega_1}^{\text{mix}}\right]\\\nonumber
&\left\{\eta_{\varepsilon}\left[\mathcal{I}_b-\mathcal{I}_c\right]+\eta_{\varepsilon-\omega_1}\left[\mathcal{I}_d-\mathcal{I}_e\right]+2i\eta_{\varepsilon-\omega_1}\left[\mathcal{I}_f+\mathcal{I}_g-\mathcal{I}_h\right]\right\};
\end{flalign}
\vspace{-7mm}
\begin{flalign}\label{2CurrentDiagaram-Elements3}\nonumber
&\mathcal{A}_3=
\left[f_{\text{R}}^{\varepsilon}-f_{\text{R}}^{\varepsilon-\Omega}\right]
[N_{\Omega}^{\text{ph}}-N_{\Omega}^{\text{R}}]\eta_{\varepsilon-\Omega}\mathcal{I}_j\\
&\hspace{-0.5mm}+\hspace{-0.5mm}\left[f_{\text{L}}^{\varepsilon}-f_{\text{L}}^{\varepsilon-\Omega}\right]
[N_{\Omega}^{\text{ph}}-N_{\Omega}^{\text{L}}]
\left[\eta_{\varepsilon}\mathcal{I}_i-2i\eta_{\varepsilon-\Omega}\mathcal{I}_o\right]\\\nonumber
&\hspace{-0.5mm}-\hspace{-0.5mm}
\left[f_{\text{R}}^{\varepsilon}-f_{\text{L}}^{\varepsilon-\Omega}\right][N_{\Omega}^{\text{ph}}-N_{\Omega}^{\text{mix}}]\left[\eta_{\varepsilon}\mathcal{I}_i+\eta_{\varepsilon-\Omega}\mathcal{I}_j+2i\eta_{\varepsilon-\Omega}\mathcal{I}_k\right].
\end{flalign}
\end{subequations}
The parameter $\eta_{\varepsilon}$  equals $-e$ for the electric current and $\varepsilon-\mu_\text{L}$ for the thermal current. The temperature entering the Fermi-Dirac distribution function of the left (right) lead is always $T_{L}$ ($T_{R}$). The Bose-Einstein distribution functions $N^{j}$ can, on the other hand, depend on the temperature of either the leads  $j=\text{L/R}$ or the phonons  $N^{\text{ph}}$. In addition, we use the shorthand notation $N_{\omega}^{\text{mix}}=N\left[{(\omega-\varepsilon+\mu_L)}/{T_{\text{L}}}+{(\varepsilon-\mu_R)}/{T_{\text{R}}}\right]$, which arises from the identity $f_{\varepsilon}^{R}(f_{\varepsilon-\omega}^{L}-1)=N_{\omega}^{\text{mix}}(f_{\varepsilon}^{R}-f_{\varepsilon-\omega}^{L})$.

The electric and thermal conductances of free electrons are connected through the Wiedemann-Franz law, which stems from the common origin of the two quantities. Each electron flowing between the two electrodes carries its charge and free energy. The equivalence between the currents and the Wiedemann-Franz law breaks in the presence of interactions~\cite{Raimondi2004,Aleiner2005,Smith2005,Michaeli}.  Specifically, an electron that is backscattered after absorbing (emitting) phonons increases (reduces) the thermal current from the subsystem to the lead of its origin. By contrast, the energy in which electrons enter the lead is irrelevant for the electric current. Therefore, the diagrams in Fig.~\ref{fig.3}, accounting for such processes, do not contribute to the electric conductance. Importantly, the electric current in response to a temperature difference between the leads \textit{is} sensitive to backscattering by phonons. The driving force of the current in this case is proportional to the electron's energy, which changes after the scattering event. Another dissimilarity between the currents is in charge conservation: the total electric current flowing out of the left lead must reach the right one, and $J_{\text{L}}+J_{\text{R}}=0$, while  heat is also transferred to the phonon baths.

In summary, we obtained an approximation for the electric and heat currents in a two-terminal setup through an interacting finite region.  This work aimed to find a tractable expression for the currents in cases where the exact GF cannot be found, but straightforward perturbation theory does not capture the physical picture accurately. Specifically, we started from the general MWF~\cite{Meir}. We derived expressions for the currents that sum over the infinite subseries of contributions containing the lowest order corrections to the self-energy. Such a simplification is known as the GW approximation, which is widely used in studies of equilibrium properties of molecules and solids. Our work provides a way of extending these calculations to obtain also transport coefficients. In particular, we expect it to be applicable for various mesoscopic and nanoscopic systems such as molecular junctions, nanotubes, or nanoribbons. Finally, we hope this work serves as a starting point for finding a self-consistent solution for the current in the presence of interactions.

\textbf{Acknowledgments}: This work was supported by the Minerva foundation with funding from the Federal German Ministry for Education and Research.

\begin{small}

\end{small}

\end{document}

\title{Landauer formula for interacting systems: a consistent non-perturbative approximation - Supplementary Material}

\author{Dan Klein}
\affiliation{Department of Condensed Matter Physics, Weizmann Institute of Science, Rehovot 76100, Israel}
\author{Karen Michaeli}
\affiliation{Department of Condensed Matter Physics, Weizmann Institute of Science, Rehovot 76100, Israel}

\maketitle
%\date{\today}

\section{Explicit expression for the current}

The main results of our paper, Eqs.~(8) and~(10), are written in terms of the diagrams shown in Fig.~2. For completeness, we give here the explicit expression for each of the diagrams:
\begin{subequations}
\begin{flalign}\tag{S1a}
\Re\mathcal{I}_a=\left[\Gamma^{L}\cdot G^{R}\cdot \Gamma^{R}\cdot  G^{A}\right]_{\mathbf{n},\mathbf{n}}^{\varepsilon};&&
\end{flalign}

\begin{flalign}\tag{S1b}
\Re\mathcal{I}_b=&\left[g^R\cdot \Gamma^{R}\cdot g^A\cdot \Gamma^{L}\cdot g^R-
g^A\cdot \Gamma^{L}\cdot g^R\cdot \Gamma^{R}\cdot g^A
\right]_{\mathbf{n}',\mathbf{n}}^{\varepsilon}\left[D^{R}(\omega_1)-D^{A}(\omega_1)\right]_{\mathbf{n},\mathbf{p};\mathbf{p}',\mathbf{n}'}
\left[G^{R}\cdot\Gamma^{L}\cdot G^{A}\right]_{\mathbf{p},\mathbf{p}'}^{\varepsilon-\omega_1};&&
\end{flalign}

\begin{flalign}\tag{S1c}
\Re\mathcal{I}_c=&\left[g^R\cdot \Gamma^{R}\cdot g^A\cdot \Gamma^{L}\cdot g^R-
g^A\cdot \Gamma^{L}\cdot g^R\cdot \Gamma^{R}\cdot g^A
\right]_{\mathbf{n}',\mathbf{n}}^{\varepsilon}\left[D^{R}(\omega_1)-D^{A}(\omega_1)\right]_{\mathbf{n},\mathbf{p};\mathbf{p}',\mathbf{n}'}
\left[g^{R}\cdot\Gamma^{L}\cdot g^{A}\right]_{\mathbf{p},\mathbf{p}'}^{\varepsilon-\omega_1};
&&
\end{flalign}

\begin{flalign}\tag{S1d}
\Re\mathcal{I}_d=&\left[g^R\cdot \Gamma^{R}\cdot g^A\cdot \Gamma^{L}\cdot g^R-
g^A\cdot \Gamma^{L}\cdot g^R\cdot \Gamma^{R}\cdot g^A
\right]_{\mathbf{n}',\mathbf{n}}^{\varepsilon-\omega_1}\left[D^{R}(\omega_1)-D^{A}(\omega_1)\right]_{\mathbf{p}',\mathbf{n}';\mathbf{n},\mathbf{p}}
\left[G^{R}\cdot\Gamma^{R}\cdot G^{A}\right]_{\mathbf{p},\mathbf{p}'}^{\varepsilon};
&&\end{flalign}

\begin{flalign}\tag{S1e}
\Re\mathcal{I}_e=&\left[g^R\cdot \Gamma^{R}\cdot g^A\cdot \Gamma^{L}\cdot g^R-
g^A\cdot \Gamma^{L}\cdot g^R\cdot \Gamma^{R}\cdot g^A
\right]_{\mathbf{n}',\mathbf{n}}^{\varepsilon-\omega_1}\left[D^{R}(\omega_1)-D^{A}(\omega_1)\right]_{\mathbf{p}',\mathbf{n}';\mathbf{n},\mathbf{p}}
\left[g^{R}\cdot\Gamma^{R}\cdot g^{A}\right]_{\mathbf{p},\mathbf{p}'}^{\varepsilon};
&&
\end{flalign}

\begin{flalign}\tag{S1f}
\Re\mathcal{I}_f= \left[ g^A\cdot \Gamma^{L}\cdot g^R
\right]_{\mathbf{n}',\mathbf{n}}^{\varepsilon-\omega_1}\left[D^{R}(\omega_1)-D^{A}(\omega_1)\right]_{\mathbf{p}',\mathbf{n}';\mathbf{n},\mathbf{p}}
\left[G^{R}\cdot\Gamma^{R}\cdot G^{A}\right]_{\mathbf{p},\mathbf{p}'}^{\varepsilon};
&&
\end{flalign}

\begin{flalign}\tag{S1g}
\Re\mathcal{I}_g= \left[ G^A\cdot \Gamma^{L}\cdot G^R
\right]_{\mathbf{n}',\mathbf{n}}^{\varepsilon-\omega_1}\left[D^{R}(\omega_1)-D^{A}(\omega_1)\right]_{\mathbf{p}',\mathbf{n}';\mathbf{n},\mathbf{p}}
\left[g^{R}\cdot\Gamma^{R}\cdot g^{A}\right]_{\mathbf{p},\mathbf{p}'}^{\varepsilon};
&&
\end{flalign}

\begin{flalign}\tag{S1h}
\Re\mathcal{I}_h= \left[ g^A\cdot \Gamma^{L}\cdot g^R
\right]_{\mathbf{n}',\mathbf{n}}^{\varepsilon-\omega_1}\left[D^{R}(\omega_1)-D^{A}(\omega_1)\right]_{\mathbf{p}',\mathbf{n}';\mathbf{n},\mathbf{p}}
\left[g^{R}\cdot\Gamma^{R}\cdot g^{A}\right]_{\mathbf{p},\mathbf{p}'}^{\varepsilon};&&
\end{flalign}

\begin{flalign}\tag{S1i}
\Re\mathcal{I}_i=&\left[g^R\cdot \Gamma^{R}\cdot g^A\cdot \Gamma^{L}\cdot g^R-
g^A\cdot \Gamma^{L}\cdot g^R\cdot \Gamma^{R}\cdot g^A
\right]_{\mathbf{n}',\mathbf{n}}^{\varepsilon}\left[D^{R}(\omega_1)-D^{A}(\omega_1)\right]_{\mathbf{n},\mathbf{p};\mathbf{p}',\mathbf{n}'}\\\nonumber
&\hspace{10mm}\times
G_{\mathbf{p},\mathbf{m}}^{R}(\varepsilon-\omega_1)
\left[D^{R}(\omega_2)-D^{A}(\omega_2)\right]_{\mathbf{m},\mathbf{q};\mathbf{q}',\mathbf{m}'}
\left[g^{R}\cdot\Gamma^{L}\cdot g^{A}\right]_{\mathbf{q},\mathbf{q}'}^{\varepsilon-\omega_1-\omega_2}
G_{\mathbf{m}',\mathbf{p}'}^{A}(\varepsilon-\omega_1);
&&
\end{flalign}

\begin{flalign}\tag{S1j}
\Re\mathcal{I}_j=&\left[g^R\cdot \Gamma^{R}\cdot g^A\cdot \Gamma^{L}\cdot g^R-
g^A\cdot \Gamma^{L}\cdot g^R\cdot \Gamma^{R}\cdot g^A
\right]_{\mathbf{n}',\mathbf{n}}^{\varepsilon-\omega_1-\omega_2}\left[D^{R}(\omega_2)-D^{A}(\omega_2)\right]_{\mathbf{p}',\mathbf{n}';\mathbf{n},\mathbf{p}}\\\nonumber
&\hspace{10mm}\times
G_{\mathbf{p},\mathbf{m}}^{R}(\varepsilon-\omega_1)
\left[D^{R}(\omega_1)-D^{A}(\omega_1)\right]_{\mathbf{q}',\mathbf{m}';\mathbf{m},\mathbf{q}}
\left[g^{R}\cdot\Gamma^{R}\cdot g^{A}\right]_{\mathbf{q},\mathbf{q}'}^{\varepsilon}
G_{\mathbf{m}',\mathbf{p}'}^{A}(\varepsilon-\omega_1);
&&
\end{flalign}

\begin{flalign}\tag{S1k}
\Re\mathcal{I}_k=& \left[g^A\cdot \Gamma^{L}\cdot g^R
\right]_{\mathbf{n}',\mathbf{n}}^{\varepsilon-\omega_1-\omega_2}\left[D^{R}(\omega_2)-D^{A}(\omega_2)\right]_{\mathbf{p}',\mathbf{n}';\mathbf{n},\mathbf{p}}\\\nonumber
&\hspace{10mm}\times
G_{\mathbf{p},\mathbf{m}}^{R}(\varepsilon-\omega_1)
\left[D^{R}(\omega_1)-D^{A}(\omega_1)\right]_{\mathbf{q}',\mathbf{m}';\mathbf{m},\mathbf{q}}
\left[g^{R}\cdot\Gamma^{R}\cdot g^{A}\right]_{\mathbf{q},\mathbf{q}'}^{\varepsilon}
G_{\mathbf{m}',\mathbf{p}'}^{A}(\varepsilon-\omega_1);
&&
\end{flalign}

\end{subequations}
In all of the above equations, we sum over repeated indices and denote the product of matrices by a dot, $[A\cdot B]_{\mathbf{n},\mathbf{n}'}^{\varepsilon}=A_{\mathbf{n},\mathbf{p}}(\varepsilon)B_{\mathbf{p},\mathbf{n}'}(\varepsilon)=\sum_{\mathbf{p}}A_{\mathbf{n},\mathbf{p}}(\varepsilon)B_{\mathbf{p},\mathbf{n}'}(\varepsilon)$.

Similarly, we can write the expressions for the additional diagrams in Fig.~4
\begin{subequations}

\begin{flalign}\tag{S2a}
\Re\mathcal{I}_l= \left[g^A\cdot \Gamma^{L}\cdot g^R
\right]_{\mathbf{n}',\mathbf{n}}^{\varepsilon-\omega_1}\left[D^{R}(\omega_1)-D^{A}(\omega_1)\right]_{\mathbf{p}',\mathbf{n}';\mathbf{n},\mathbf{p}}
\left[G^{R}\cdot\Gamma^{L}\cdot G^{A}\right]_{\mathbf{p},\mathbf{p}'}^{\varepsilon};
&&
\end{flalign}

\begin{flalign}\tag{S2b}
\Re\mathcal{I}_m= \left[G^A\cdot \Gamma^{L}\cdot G^R
\right]_{\mathbf{n}',\mathbf{n}}^{\varepsilon-\omega_1}\left[D^{R}(\omega_1)-D^{A}(\omega_1)\right]_{\mathbf{p}',\mathbf{n}';\mathbf{n},\mathbf{p}}
\left[g^{R}\cdot\Gamma^{L}\cdot g^{A}\right]_{\mathbf{p},\mathbf{p}'}^{\varepsilon};
&&
\end{flalign}

\begin{flalign}\tag{S2c}
\Re\mathcal{I}_n= \left[g^A\cdot \Gamma^{L}\cdot g^R
\right]_{\mathbf{n}',\mathbf{n}}^{\varepsilon-\omega_1}\left[D^{R}(\omega_1)-D^{A}(\omega_1)\right]_{\mathbf{p}',\mathbf{n}';\mathbf{n},\mathbf{p}}
\left[g^{R}\cdot\Gamma^{L}\cdot g^{A}\right]_{\mathbf{p},\mathbf{p}'}^{\varepsilon};&&
\end{flalign}

\begin{flalign}\tag{S2d}
\Re\mathcal{I}_o=& \left[g^A\cdot \Gamma^{L}\cdot g^R
\right]_{\mathbf{n}',\mathbf{n}}^{\varepsilon-\omega_1-\omega_2}\left[D^{R}(\omega_2)-D^{A}(\omega_2)\right]_{\mathbf{n},\mathbf{p};\mathbf{p}',\mathbf{n}'}\\\nonumber
&\hspace{10mm}\times
G_{\mathbf{p},\mathbf{m}}^{R}(\varepsilon-\omega_1)
\left[D^{R}(\omega_1)-D^{A}(\omega_1)\right]_{\mathbf{q}',\mathbf{m}';\mathbf{m},\mathbf{q}}
\left[g^{R}\cdot\Gamma^{L}\cdot g^{A}\right]_{\mathbf{m},\mathbf{m}'}^{\varepsilon}
G_{\mathbf{m}',\mathbf{p}'}^{A}(\varepsilon-\omega_1);
&&
\end{flalign}

\end{subequations}

\section{Current conservation}

This section aims to show that our expression for the electric current does not violate charge conservation. Below, we show that currents flowing between the subsystem and the left and right leads satisfy  $J_{\text{L}}+J_{\text{R}}=0$. To generate the expression for $J_{\text{R}}$, we replace the left and right leads in Eq.~(9). In preparation to add the two currents, we shift the fermion's frequency under the integral in $J_{\text{R}}$ as follows $\varepsilon\rightarrow \varepsilon+\omega$ (or $\varepsilon\rightarrow \varepsilon+\omega_1+\omega_2$). Similarly, we substitute $\omega_i\rightarrow-\omega_i$ for all bosons' frequencies. Next, we write the sum of the two currents as   
\begin{align}\tag{S3}
J_{\text{L}}+J_{\text{R}}=J^{(0)}+J^{(1)}+J^{(2)}.
\end{align}
Here, $J^{(i)}$ is the contribution to the currents from diagrams in which $i$ phononic propagators explicitly appear. Namely, $J^{\left( 0 \right)}$ contains $\mathcal{I}_{a}$ from Fig.~2 of the main text and its equivalent contribution to the current in the right lead. The term $J^{(1)}$ contains diagrams $\mathcal{I}_{b}$-$\mathcal{I}_{h}$, and the remaining diagrams enter $J^{(2)}$. Next, we simplify the expressions for $J^{(1)}$ and $J^{(2)}$ by applying the identity given by Eq.~(4) in the main text on the bare Green's function. In addition, we use the relation $N^{\text{ph}}_{\omega_{1}} N^{\text{ph}}_{\omega_{2}} = N^{\text{ph}}_{\omega_{1} + \omega_{2}} \left[ N^{\text{ph}}_{\omega_{1}} - N^{\text{ph}}_{-\omega_{2}}\right]$.  Thus, the different contributions to $J_{\text{L}}+J_{\text{R}}$ can be written as
\begin{subequations}
\begin{flalign}\tag{S4a}
J^{(0)}=-\frac{e}{\hbar}\int{\mathrm{d\varepsilon}}\left[f_{\text{L}}^{\varepsilon}-f_{\text{R}}^{\varepsilon}\right]
\left[\Gamma^{L}\cdot G^{R}\cdot \Gamma^{R}\cdot  G^{A}-\Gamma^{R}\cdot G^{R}\cdot \Gamma^{L}\cdot  G^{A}\right]_{\mathbf{n},\mathbf{n}}^{\varepsilon};
&&
\end{flalign}
\begin{flalign}\tag{S4b}
&J^{(1)}=i\frac{e}{\hbar}\int\frac{\mathrm{d\varepsilon d\omega_1}}{2\pi}\left[f_{\text{R}}^{\varepsilon}-f_{\text{L}}^{\varepsilon-\omega_1}\right]\left[N_{\omega_1}^{\text{ph}}-N_{\omega_1+eV}^{\text{ph}}\right]
\left[D^{R}(\omega_1)-D^{A}(\omega_1)\right]_{\mathbf{n},\mathbf{p};\mathbf{p}',\mathbf{n}'}
\\\nonumber
&\left\{\left[g^R\cdot \Gamma^{L}\cdot g^A\right]_{\mathbf{p},\mathbf{p}'}^{\varepsilon-\omega_1}\left[G^{R}\cdot\Gamma^{R}\cdot G^{A}-G^A\cdot \Gamma^{R}\cdot G^R\right]_{\mathbf{n}',\mathbf{n}}^{\varepsilon}-\left[g^R\cdot \Gamma^{R}\cdot g^A\right]_{\mathbf{n}',\mathbf{n}}^{\varepsilon}\left[G^{R}\cdot\Gamma^{L}\cdot G^{A}-G^A\cdot \Gamma^{L}\cdot G^R\right]_{\mathbf{p},\mathbf{p}'}^{\varepsilon-\omega_1}
\right\};
&&
\end{flalign}
\begin{flalign}\tag{S4c}
&J^{(2)}=\frac{e}{\hbar}\int\frac{\mathrm{d\varepsilon d\omega_1d\omega_2}}{(2\pi)^2}
\left[f_{\text{R}}^{\varepsilon}-f_{\text{L}}^{\varepsilon-\omega_1-\omega_2}\right]
\left[N_{\omega_2}^{\text{ph}}-N_{-\omega_1}^{\text{ph}}\right]\left[N_{\omega_1+\omega_2}^{\text{ph}}-N_{\omega_1+\omega_2+eV}^{\text{ph}}\right]\\\nonumber
&\times
\left[D^{R}(\omega_1)-D^{A}(\omega_1)\right]_{\mathbf{n},\mathbf{p};\mathbf{p}',\mathbf{n}'}
\left[D^{R}(\omega_2)-D^{A}(\omega_2)\right]_{\mathbf{m},\mathbf{q};\mathbf{q}',\mathbf{m}'}
\left[g^R\cdot \Gamma^{R}\cdot g^A\right]_{\mathbf{n}',\mathbf{n}}^{\varepsilon}\\\nonumber
&\times
\left[G_{\mathbf{p},\mathbf{m}}^{R}(\varepsilon-\omega_1)
G_{\mathbf{m}',\mathbf{p}'}^{A}(\varepsilon-\omega_1)-G_{\mathbf{p},\mathbf{m}}^{A}(\varepsilon-\omega_1)
G_{\mathbf{m}',\mathbf{p}'}^{R}(\varepsilon-\omega_1)
\right]
\left[g^R\cdot \Gamma^{L}\cdot g^A\right]_{\mathbf{q},\mathbf{q}'}^{\varepsilon-\omega_1-\omega_2}.
&&
\end{flalign}
\end{subequations}
Although each term seems to have a different number of phonon propagators, it comprises an infinite sum of corrections of all orders in the interaction. This infinite sum is generated by the fully dressed Green's functions $G$. Consequently, each term is generally not zero, and current conservation is restored only upon summing all terms.   

Our next step is to sum $J^{(0)}$ and $J^{(1)}$. For this purpose, we use Eq.~(4) in the main text to rewrite $J^{(0)}$ as a function of the self-energy
\begin{align}\tag{S5}
&J^{(0)}=-i\frac{e}{2\hbar}\int{\mathrm{d\varepsilon}}\left[f_{\text{L}}^{\varepsilon}-f_{\text{R}}^{\varepsilon}\right]
\left[ G^{A}\cdot (\Gamma^{R}-\Gamma^{L})\cdot  G^{R}-G^{R}\cdot (\Gamma^{R}-\Gamma^{L})\cdot  G^{A}\right]_{\mathbf{n}',\mathbf{n}}^{\varepsilon}\left[\Sigma^R-\Sigma^A\right]_{\mathbf{n},\mathbf{n}'}^{\varepsilon}
\\\nonumber
&=i\frac{e}{2\hbar}\int\frac{\mathrm{d\varepsilon d\omega_1}}{2\pi}
\left[f_{\text{R}}^{\varepsilon}-f_{\text{L}}^{\varepsilon-\omega_1}\right]\left[N_{\omega_1}^{\text{ph}}-N_{\omega_1+eV}^{\text{ph}}\right]
\left[D^{R}(\omega_1)-D^{A}(\omega_1)\right]_{\mathbf{n},\mathbf{p};\mathbf{p}',\mathbf{n}'}
\left\{\left[g^R\cdot \Gamma^{R}\cdot g^A\right]_{\mathbf{n}',\mathbf{n}}^{\varepsilon}
\left[G^{A}\cdot(\Gamma^{R}-\Gamma^{L})\cdot  G^{R}\right.\right.\\\nonumber
&\left.\left.
-G^{R}\cdot(\Gamma^{R}-\Gamma^{L})\cdot  G^{A}\right]_{\mathbf{p},\mathbf{p}'}^{\varepsilon-\omega_1}
+
\left[ G^{A}\cdot(\Gamma^{R}-\Gamma^{L})\cdot  G^{R}-G^{R}\cdot(\Gamma^{R}-\Gamma^{L})\cdot G^{A}\right]_{\mathbf{n}',\mathbf{n}}^{\varepsilon}
\left[g^R\cdot\Gamma^{L}\cdot g^A\right]_{\mathbf{p},\mathbf{p}'}^{\varepsilon-\omega_1} \right\}.
\end{align}
We, thus, brought $J^{(0)}$ into a form similar to $J^{(1)}$ and their sum yields
\begin{align}\tag{S6}
J^{(0)}+J^{(1)}&=
i\frac{e}{2\hbar}\int\frac{\mathrm{d\varepsilon d\omega_1}}{2\pi}
\left[f_{\text{R}}^{\varepsilon}-f_{\text{L}}^{\varepsilon-\omega_1}\right]\left[N_{\omega_1}^{\text{ph}}-N_{\omega_1+eV}^{\text{ph}}\right]
\left[D^{R}(\omega_1)-D^{A}(\omega_1)\right]_{\mathbf{n},\mathbf{p};\mathbf{p}',\mathbf{n}'}\\\nonumber
&\left\{\left[g^R\cdot \Gamma^{R}\cdot g^A\right]_{\mathbf{n}',\mathbf{n}}^{\varepsilon}
\left[G^{A}\cdot (\Gamma^{R}+\Gamma^{L})\cdot  G^{R}-G^{R}\cdot (\Gamma^{R}+\Gamma^{L})\cdot  G^{A}\right]_{\mathbf{p},\mathbf{p}'}^{\varepsilon-\omega_1}
\right.\\\nonumber&\left.\hspace{40mm}+
\left[G^{R}\cdot (\Gamma^{R}+\Gamma^{L})\cdot  G^{A}-G^{A}\cdot (\Gamma^{R}+\Gamma^{L})\cdot  G^{R}\right]_{\mathbf{n}',\mathbf{n}}^{\varepsilon}
\left[g^R\cdot \Gamma^{L}\cdot g^A\right]_{\mathbf{p},\mathbf{p}'}^{\varepsilon-\omega_1}
\right\}.
\end{align}
Finally, we apply again Eq.~(4) to rewrite the above equation
\begin{align}\tag{S7}
&J^{(0)}+J^{(1)}=\frac{e}{2\hbar}\int\frac{\mathrm{d\varepsilon d\omega_1}}{2\pi}
\left[f_{\text{R}}^{\varepsilon}-f_{\text{L}}^{\varepsilon-\omega_1}\right]\left[N_{\omega_1}^{\text{ph}}-N_{\omega_1+eV}^{\text{ph}}\right]
\left[D^{R}(\omega_1)-D^{A}(\omega_1)\right]_{\mathbf{n},\mathbf{p};\mathbf{p}',\mathbf{n}'}\\\nonumber
&\times
\left\{\left[g^R\cdot \Gamma^{R}\cdot g^A\right]_{\mathbf{n}',\mathbf{n}}^{\varepsilon}
\left[G_{\mathbf{p},\mathbf{m}}^{A}(\varepsilon-\omega_1)G_{\mathbf{m}',\mathbf{p}'}^{R}(\varepsilon-\omega_1)-
G_{\mathbf{p},\mathbf{m}}^{R}(\varepsilon-\omega_1)G_{\mathbf{m}',\mathbf{p}'}^{A}(\varepsilon-\omega_1)\right]
\left[\Sigma^{R}(\varepsilon-\omega_1)-\Sigma^{A}(\varepsilon-\omega_1)\right]_{\mathbf{m},\mathbf{m}'}
\right.\\\nonumber
&\left.+\left[G_{\mathbf{n}',\mathbf{m}}^{R}(\varepsilon)G_{\mathbf{m}',\mathbf{n}}^{A}(\varepsilon)-
G_{\mathbf{n}',\mathbf{m}}^{A}(\varepsilon)G_{\mathbf{m}',\mathbf{n}}^{R}(\varepsilon)\right]
\left[\Sigma^{R}(\varepsilon)-\Sigma^{A}(\varepsilon)\right]_{\mathbf{m},\mathbf{m}'}
\left[g^R\cdot \Gamma^{L}\cdot g^A\right]_{\mathbf{p},\mathbf{p}'}^{\varepsilon-\omega_1}
\right\}\\\nonumber
&=\frac{e}{\hbar}\int\frac{\mathrm{d\varepsilon d\omega_1 d\omega_2}}{(2\pi)^2}
\left[g^R\cdot \Gamma^{R}\cdot g^A\right]_{\mathbf{n}',\mathbf{n}}^{\varepsilon}
\left[G_{\mathbf{p},\mathbf{m}}^{A}(\varepsilon-\omega_1)G_{\mathbf{m}',\mathbf{p}'}^{R}(\varepsilon-\omega_1)-
G_{\mathbf{p},\mathbf{m}}^{R}(\varepsilon-\omega_1)G_{\mathbf{m}',\mathbf{p}'}^{A}(\varepsilon-\omega_1)\right]
\left[g^R\cdot \Gamma^{L}\cdot g^A\right]_{\mathbf{q},\mathbf{q}'}^{\varepsilon-\omega_1-\omega_2}\\\nonumber
&\times
\left[f_{\text{R}}^{\varepsilon}-f_{\text{L}}^{\varepsilon-\omega_1-\omega_2}\right]\left[N_{\omega_1+\omega_2}^{\text{ph}}-N_{\omega_1+\omega_2+eV}^{\text{ph}}\right]
\left[N_{\omega_2}^{\text{ph}}-N_{-\omega_1}^{\text{ph}}\right]
\left[D^{R}(\omega_1)-D^{A}(\omega_1)\right]_{\mathbf{n},\mathbf{p};\mathbf{p}',\mathbf{n}'}
\left[D^{R}(\omega_2)-D^{A}(\omega_2)\right]_{\mathbf{m},\mathbf{q};\mathbf{q}',\mathbf{m}'}
\end{align}
We found that $J^{(0)}+J^{(1)}=-J^{(2)}$ and, hence, our expression for the current conserves charge.

\section{The electric current in the presence of electron-electron interactions}

We devote this section to modifying the expression for the current to include renormalization of the Boson properties by the interactions. Such a scenario is relevant for small bosonic baths and electron-electron interactions, where the renormalization is significant. We start by considering a system coupled to a phonon bath, i.e., we return to the example given in the main text when only a voltage bias is applied to the system. We derived the current for systems where the renormalization of the phonons properties is negligible. Consequently, the Boson propagator used in the main text is simply $D^{R/A}(\vec{k},\omega)\propto[(\omega-\omega_{\vec{k}}\pm i\delta)^{-1}-(\omega+\omega_{\vec{k}}\pm i\delta)^{-1}]$. By contrast, here we are interested in systems for which this expression is no longer valid, and the dressed propagator becomes 
\begin{align}\tag{S8}
\left[U^{R,A}\right]^{-1}=\left[D^{R,A}\right]^{-1}-\Pi^{R/A}.
\end{align}
The lesser and greater components of the Green's function are
\begin{align}\tag{S9}
U^{<,>}(\vec{k},\omega)=U^R\cdot\Pi^{<,>}\cdot U^{A}.
\end{align}

To proceed with the derivation of the current, we need to specify the form of the phonons self-energy (polarization operator) $\Pi$. Here, we derived the phonons' self-energy within the random phase approximation (RPA)
\begin{flalign}\tag{S10}
\Pi_{\mathbf{n},\mathbf{p};\mathbf{p}',\mathbf{n}'}^{R,A}(\omega)=-\sum_{j=L/R}\int\frac{\mathrm{d\varepsilon}}{2\pi}
\left\{f_{j}(\varepsilon)\left[g^{R}\Gamma^{j}g^{A}\right]_{\mathbf{n}',\mathbf{n}}^{\varepsilon}g_{\mathbf{p},\mathbf{p}'}^{A,R}(\varepsilon-\omega)+
f_{j}(\varepsilon-\omega)g_{\mathbf{n}',\mathbf{n}}^{R,A}(\varepsilon)\left[g^{R}\Gamma^{j}g^{A}\right]_{\mathbf{p},\mathbf{p}'}^{\varepsilon-\omega}\right\}.
\end{flalign}
We use the expression for the retard and advanced components of the self-energy to introduce a new quantity
\begin{flalign}\tag{S11}\label{PiIndices}
\Pi_{\mathbf{n},\mathbf{p};\mathbf{p}',\mathbf{n}'}^{R}(\omega)-\Pi_{\mathbf{n},\mathbf{p};\mathbf{p}',\mathbf{n}'}^{A}(\omega)&=-i\sum_{j,\ell=L/R}\int\frac{\mathrm{d\varepsilon}}{2\pi}
\left[f_{j}(\varepsilon)-f_{\ell}(\varepsilon-\omega)\right](g^{R}\Gamma^{j}g^{A})_{\mathbf{n}',\mathbf{n}}^{\varepsilon}(g^{R}\Gamma^{\ell}g^{A})_{\mathbf{p},\mathbf{p}'}^{\varepsilon-\omega}\\\nonumber
&\equiv\sum_{j,\ell=R/L}\left[
\Pi_{j,\ell}^{R}(\omega)-\Pi_{j,\ell}^{A}(\omega)\right]_{\mathbf{n},\mathbf{p};\mathbf{p}',\mathbf{n}'}.
\end{flalign}
The indices $j$ and $\ell$ in $\Pi_{j,\ell}^{R}(\omega)-\Pi_{j,\ell}^{A}(\omega)$ denote the leads through which the Green's function in the polarization operator pass and relax into. Consequently, the lesser and greater components of $\Pi$ can be written as
\begin{subequations}
\begin{flalign}\tag{S12a}
\Pi^{<}=\sum_{j,\ell=R/L}N_{\omega+\mu_{\ell}-\mu_{j}}^{\text{ph}}\left[
\Pi_{j,\ell}^{R}-\Pi_{j,\ell}^{A}\right];
\end{flalign}
\begin{flalign}\tag{S12b}
\Pi^{>}=\sum_{j,\ell=R/L}\left[1+N_{\omega+\mu_{\ell}-\mu_{j}}^{\text{ph}}\right]\left[
\Pi_{j,\ell}^{R}-\Pi_{j,\ell}^{A}\right].
\end{flalign}
Finally, similar to Eq.~\eqref{PiIndices}, we define $U_{j,\ell}^{R}(\omega)-U_{j,\ell}^{A}(\omega)$ 
\begin{align}\tag{S13}
U^{R}-U^{A}=\sum_{j,\ell=L/R}\left[U^R\cdot(\Pi_{j,\ell}^{R}-\Pi_{j,\ell}^{A})\cdot U^{A}\right]\equiv\sum_{j,\ell=L/R}\left[U_{j,\ell}^{R}-U_{j,\ell}^{A}\right].
\end{align}
\end{subequations}

We can now write the expression for the current using the diagrammatic representation appearing in the main text, similar to Eq.~(9). In cases where the renormalization of the boson modes is significant, the double wiggly line entering the self-energy (Fig.~1) represents the full phonon propagator (S8) and (S9). The single wiggly line in Figs.~2 and~3 denotes $U_{j,\ell}^{R}-U_{j,\ell}^{A}$. We demonstrate the changes in  $\mathcal{I}_{\alpha}$ on $\mathcal{I}_{b}$ (Eq.~(S1b)) that becomes
\begin{flalign}\tag{S14}
\Re\mathcal{I}_b=& \sum_{j,\ell=L/R}\left[g^R\cdot \Gamma^{R}\cdot g^A\cdot \Gamma^{L}\cdot g^R-
g^A\cdot \Gamma^{L}\cdot g^R\cdot \Gamma^{R}\cdot g^A
\right]_{\mathbf{n}',\mathbf{n}}^{\varepsilon}\left[U_{j,\ell}^{R}(\omega_1)-U_{j,\ell}^{A}(\omega_1)\right]_{\mathbf{n},\mathbf{p};\mathbf{p}',\mathbf{n}'}
\left[G^{R}\cdot\Gamma^{L}\cdot G^{A}\right]_{\mathbf{p},\mathbf{p}'}^{\varepsilon-\omega_1}.
\end{flalign}
Then, we can write the expression for the current with the modified $\mathcal{I}_{\alpha}$
\begin{align}\tag{S15}
&J_{\text{L}}=-\frac{e}{\hbar}\int{\mathrm{d\varepsilon}}\Re\mathcal{I}_a\left[f_{\text{L}}^{\varepsilon}-f_{\text{R}}^{\varepsilon}\right]\\\nonumber
&
+\frac{e}{2\hbar}\int\frac{\mathrm{d\varepsilon d\omega_1}}{2\pi}\sum_{j_1,\ell_1=L/R}
\Re\left[\mathcal{I}_b-\mathcal{I}_c+\mathcal{I}_d-\mathcal{I}_e+2i\mathcal{I}_f+\mathcal{I}_g-\mathcal{I}_h\right]
\left[f_{\text{R}}^{\varepsilon}-f_{\text{L}}^{\varepsilon-\omega_1}\right]\left[N_{\omega_1+\Delta\mu_{j_1,\ell_1}}^{\text{ph}}-N_{\omega_1+eV}^{\text{ph}}\right]
\\\nonumber
&
+\frac{e}{2\hbar}\int\frac{\mathrm{d\varepsilon d\omega_1}}{2\pi}\hspace{-1mm}\sum_{j_1,\ell_1=L/R}\hspace{-1mm}
\left\{\Re\left[\mathcal{I}_b-\mathcal{I}_c-2i\mathcal{I}_l+\mathcal{I}_m-\mathcal{I}_n\right]
\left[f_{\text{L}}^{\varepsilon}\hspace{-0.5mm}-\hspace{-0.5mm}f_{\text{L}}^{\varepsilon-\omega_1}\right]+\Re\left[\mathcal{I}_d-\mathcal{I}_e\right]
\left[f_{\text{R}}^{\varepsilon}-f_{\text{R}}^{\varepsilon-\omega_1}\right]\right\}\left[N_{\omega_1}^{\text{ph}}-N_{\omega_1+\Delta\mu_{j_1,\ell_1}}^{\text{ph}}\right]
\\\nonumber
&
+i\frac{e}{2\hbar}\int\frac{\mathrm{d\varepsilon d\omega_1 d\omega_2}}{(2\pi)^2}\hspace{-3mm}
\sum_{j_1,j_2,\ell_1,\ell_2=L/R}\hspace{-3mm}
\Re\left[\mathcal{I}_i+\mathcal{I}_j+2i\mathcal{I}_k\right]
\left[f_{\text{R}}^{\varepsilon}-f_{\text{L}}^{\varepsilon-\omega_1-\omega_2}\right]\left[N_{\omega_1+\omega_2+\Delta\mu_{j_1,\ell_1}+\Delta\mu_{j_2,\ell_2}}^{\text{ph}}-N_{\omega_1+\omega_2+eV}^{\text{ph}}\right]\\\nonumber
&\times
\left[N_{\omega_2+\Delta\mu_{j_2,\ell_2}}^{\text{ph}}\hspace{-0.5mm}-N_{\omega_1+\Delta\mu_{j_1,\ell_1}}^{\text{ph}}\right]\\\nonumber
&
+i\frac{e}{2\hbar}\int\frac{\mathrm{d\varepsilon d\omega_1d\omega_2}}{(2\pi)^2}
\hspace{-3mm}\sum_{j_1,j_2,\ell_1,\ell_2=L/R}\hspace{-3mm}
\left\{\Re\left[\mathcal{I}_i-2i\mathcal{I}_o\right]
\left[f_{\text{L}}^{\varepsilon}-f_{\text{L}}^{\varepsilon-\omega_1-\omega_2}\right]
+\Re\mathcal{I}_j\left[f_{\text{R}}^{\varepsilon}-f_{\text{R}}^{\varepsilon-\omega_1-\omega_2}\right]\right\}
\left[N_{\omega_2+\Delta\mu_{j_2,\ell_2}}^{\text{ph}}-N_{-\omega_1-\Delta\mu_{j_1,\ell_1}}^{\text{ph}}\right]
\\\nonumber
&\times
\left[N_{\omega_1+\omega_2}^{\text{ph}}-N_{\omega_1+\omega_2+\Delta\mu_{j_1,\ell_1}+\Delta\mu_{j_2,\ell_2}}^{\text{ph}}\right].
\end{align}
Here  $\Delta\mu_{j,\ell}=\mu_{j}-\mu_{\ell}$. Applying the same procedure as in the previous section, we verified that the current is conserved. 

So far, we discussed only coupling to boson baths; we turn now to address electron-electron interactions 
\begin{flalign}\tag{S16}
H_{\text{int}}=\sum_{n,n',\vec{m},\vec{m}',}V_{n',\vec{m}';n,\vec{m}}c_{n',\vec{m}'}^{\dag}c_{n',\vec{m}'}c_{n,\vec{m}}^{\dag}c_{n,\vec{m}}
\end{flalign}
Neglecting renormalization effects, the propagator of the field mediating the interaction is $D_{\mathbf{n},\mathbf{p};\mathbf{p}',\mathbf{n}'}^{R/A}(\vec{k},\omega)=V_{\vec{n}';\vec{n}}\delta_{\vec{n},\vec{p}}\delta_{\vec{n}',\vec{p}'}$ and $D_{\mathbf{n},\mathbf{p};\mathbf{p}',\mathbf{n}'}^{</>}(\vec{k},\omega)=0$. Accounting for self-energy corrections, we find that Eqs.~(S10)-(S13) also describe the boson propagator here. Correspondingly, Eq.~(S15) is also valid for the current through a system with dominant electron-electron interactions.